\documentclass[12pt]{iopart}

\usepackage{cite}
\usepackage{graphicx}
\usepackage{dcolumn}

\begin{document}

\title{Further analysis of the connected moments expansion}

\author{Paolo Amore \dag , Francisco M Fern\'andez \footnote[2]{Corresponding author}
 \ and Martin Rodriguez \dag}

\address{\dag\ Facultad de Ciencias, Universidad de Colima, Bernal D\'iaz del
Castillo 340, Colima, Colima, Mexico}

\ead{paolo.amore@gmail.com}

\ead{martinalexander.rv@gmail.com}

\address{\ddag\ INIFTA (UNLP, CCT La Plata-CONICET), Divisi\'on Qu\'imica Te\'orica,
Blvd. 113 S/N,  Sucursal 4, Casilla de Correo 16, 1900 La Plata,
Argentina}

\ead{fernande@quimica.unlp.edu.ar} \maketitle

\begin{abstract}
By means of simple quantum--mechanical models we show that under certain
conditions the main assumptions of the connected moments expansion (CMX) are
no longer valid. In particular we consider two--level systems, the harmonic
oscillator and the pure quartic oscillator. Although derived from such
simple models, we think that the results of this investigation may be of
utility in future applications of the approach to realistic problems. We
show that a straightforward analysis of the CMX exponential parameters may
provide a clear indication of the success of the approach.
\end{abstract}

\section{Introduction}

\label{sec:intro}

Some time ago Horn and Weinstein\cite{HW84} introduced the
$t$--expansion for the calculation of the ground--state energy of
quantum--mechanical systems. It is a Taylor expansion about $t=0$
of a generating function $E(t)$ where the coefficients are
cumulants or connected moments of the Hamiltonian operator. The
main problem posed by this approach is the extrapolation of the
$t$--power series for $t\rightarrow \infty $ in order to obtain
the ground--state energy. Horn and Weinstein\cite{HW84} and Horn
et al\cite {HKW85} proposed the use of Pad\'{e} approximants and
later Stubbins\cite {S88} tried some other extrapolation
techniques. Without doubt, the most popular extrapolation strategy
was proposed by Ciowsloski\cite{C87a} and it gives rise to the
connected--moments expansion (CMX) and leads to an expression for
the systematic calculation of the energy of the ground state.
Knowles\cite{K87} studied the CMX and derived an elegant and
compact expression for the approximants in terms of matrices built
from the cumulants.

It is well--known that the series of CMX approximants for the energy
exhibits singularities and convergence problems that limit its usefulness%
\cite{K87,MPM91,LL93,MZM94,MZMMP94,U95,AFR10}. In an attempt to
overcome those difficulties some authors proposed variants of the
CMX\cite
{MZM94,MZMMP94,U95,MFB99,FMB02,MMFB05,FMMB06,FMBC08,FCMMM10}.

The $t$--expansion, as well as the CMX and its variants, have been
tested on several simple models with varied success\cite
{S88,C87b,C87c,C87d,C87e,LW94,MFB99,MZMMP94,FMBC08,F09,AF09b,FCMMM10},
and in spite of their notorious limitations they have even been
applied to several problems of physical interest\cite
{HW84,HKW85,S88,MBM89,MPM91,MM91,MZMMP94,FMB02,MMFB05,AFR10},
including the calculation of the electronic energy of atoms and
some small molecules\cite {C87a,K87,CKSP87,YI88,NSS02}. It is
well--known that the convergence properties of the series of CMX
approximants may be considerably poorer than those of the
Rayleigh--Ritz variational method in the Krylov space and the
Lanczos algorithm\cite
{K87,MPM91,MM91,MZMMP94,MFB99,FMB02,F09,AF09b,AFR10}. However,
they can be improved by an appropriate choice of the reference
function\cite{TS11}. The main reason for still insisting on the
development of the CMX appears to be its size
consistency\cite{HW84} which those other approaches do not obey.
However, in most of the applications summarized above size
consistency is not an issue.

In this paper we test the main assumptions of the CMX by means of simple
quantum--mechanical models. In this way we expect to draw useful conclusions
that may apply to realistic problems for which such a detailed analysis is
not feasible. In section~\ref{sec:t-exp} we outline the main ideas behind
this approach. In section~\ref{sec:simple_models} we apply the CMX to $n$%
--level models. In section~\ref{sec:oscillators} we resort to the harmonic
oscillator and a nontrivial anharmonic oscillator. Finally, in section~\ref
{sec:conclusions} we discuss the results and draw conclusions.

\section{The Cumulant or $t$--expansion}

\label{sec:t-exp}

In order to facilitate present discussion, in this section we outline the
main ideas behind the $t$--expansion (or cumulant expansion) and the CMX.
The moment--generating function
\begin{equation}
Z(t)=\left\langle \varphi \right| e^{-t\hat{H}}\left| \varphi \right\rangle
=\sum_{j=0}^{\infty }\frac{(-t)^{j}}{j!}\mu _{j}  \label{eq:Z(t)}
\end{equation}
gives us the moments of the Hamiltonian operator $\hat{H}$, $\mu
_{j}=\left\langle \varphi \right| \hat{H}^{j}\left| \varphi \right\rangle $,
in the reference or trial state $\left| \varphi \right\rangle $ that we
assume to be normalized $\left\langle \varphi \right. \left| \varphi
\right\rangle =1$. The logarithmic derivative of this function
\begin{equation}
E(t)=-\frac{Z^{\prime }(t)}{Z(t)}=\frac{\left\langle \varphi \right| \hat{H}%
e^{-t\hat{H}}\left| \varphi \right\rangle }{\left\langle \varphi \right|
e^{-t\hat{H}}\left| \varphi \right\rangle }  \label{eq:E(t)}
\end{equation}
exhibits several interesting properties:

\begin{itemize}
\item  $E(t)\geq E_{0}$ for all $t$, where $E_{0}$ is the ground--state
energy.

\item  $E^{\prime }(t)\leq 0$

\item  $\lim_{t\rightarrow \infty }E(t)=E_{0}$ provided that the overlap
between $\left| \varphi \right\rangle $ and the ground state $\left| \psi
_{0}\right\rangle $ is nonzero ($\left\langle \psi _{0}\right. \left|
\varphi \right\rangle \neq 0$).
\end{itemize}

The function $E(t)$ is closely related to the cumulant function $K(t)$
defined by $Z(t)=e^{K(t)}$\cite{K62}. The formal Taylor series of $E(t)$
about $t=0$ yields the $t$--expansion (also known as cumulant or cluster
expansion):
\begin{equation}
E(t)=\sum_{j=0}^{\infty }\frac{(-t)^{j}}{j!}I_{j+1}  \label{eq:t-exp}
\end{equation}
where the cumulants $I_{j}$ (or connected moments) can be easily obtained
from the recurrence relation\cite{HW84}
\begin{equation}
I_{j+1}=\mu _{j+1}-\sum_{i=0}^{j-1}\left(
\begin{array}{c}
j \\
i
\end{array}
\right) I_{i+1}\mu _{j-i},\;j=0,1,\ldots  \label{eq:I_j}
\end{equation}
The main goal of the approach proposed by Horn and Weinstein\cite{HW84} and
Horn et al\cite{HKW85} is to find an appropriate summation method for the
cumulant expansion (\ref{eq:t-exp}) and extrapolate the resulting expression
for $t\rightarrow \infty $ to obtain $E_{0}$.

In order to carry out such extrapolation Cioslowsky\cite{C87a} proposed an
expansion in terms of exponentials

\begin{equation}
E(t)=E_{0}+\sum_{j=1}^{\infty }A_{j}\exp (-b_{j}t)  \label{eq:exp-exp}
\end{equation}
where one obtains the adjustable parameters $E_{0}$, $A_{j}$, and $b_{j}$ by
straightforwardly matching the Taylor series about $t=0$ of the left-- and
right-- hand sides. Cioslowsky\cite{C87a} showed that one can obtain $E_{0}$
without explicitly calculating the nonlinear parameters $b_{j}$. From the
properties of the Pad\'{e} approximants Knowles\cite{K87} derived an
elegant, compact and systematic expression for the calculation of $E_{0}$:
\begin{equation}
E_{0}^{(m)}=I_{1}+\left(
\begin{array}{llll}
I_{2} & I_{3} & \cdots & I_{m+1}
\end{array}
\right) \left(
\begin{array}{llll}
I_{3} & I_{4} & \cdots & I_{m+2} \\
I_{4} & I_{5} & \cdots & I_{m+3} \\
\vdots & \vdots & \ddots & \vdots \\
I_{m+2} & I_{m+3} & \cdots & I_{2m+1}
\end{array}
\right) ^{-1}\left(
\begin{array}{l}
I_{2} \\
I_{3} \\
\vdots \\
I_{m+1}
\end{array}
\right)  \label{eq:E0^(m)}
\end{equation}
The CMX will be successful provided that $\lim_{m\rightarrow
\infty }E_{0}^{(m)}=E_{0}$ if $\left\langle \psi _{0}\right.
\left| \varphi \right\rangle \neq 0$.

For simplicity, throughout this paper we assume that the
eigenfunctions $\left| \psi _{j}\right\rangle $ of the Hamiltonian
operator $\hat{H}$
\begin{equation}
\hat{H}\left| \psi _{j}\right\rangle =E_{j}\left| \psi _{j}\right\rangle
,\;E_{0}<E_{1}\leq E_{2}\leq \ldots
\end{equation}
form a complete basis set so that
\begin{equation}
\left| \varphi \right\rangle =\sum_{j=0}^{\infty }c_{j}\left| \psi
_{j}\right\rangle ,\;c_{j}=\left\langle \psi _{j}\right. \left| \varphi
\right\rangle
\end{equation}
provided that $\left\langle \psi _{i}\right. \left| \psi _{j}\right\rangle
=\delta _{ij}$. Under such conditions we can write
\begin{equation}
E(t)=\frac{\sum_{j=0}^{\infty }|c_{j}|^{2}E_{j}e^{-tE_{j}}}{%
\sum_{j=0}^{\infty }|c_{j}|^{2}e^{-tE_{j}}}  \label{eq:E(t)_cj}
\end{equation}
which clearly shows that $\lim_{t\rightarrow \infty
}E(t)=\min_{\{j\}}\{E_{j},c_{j}\neq 0\}$.

We may say that the main assumption in the CMX is that equation
(\ref {eq:exp-exp}) is valid provided that $c_{0}\neq 0$. In such
a case the ground--state energy is given by the limit of the CMX
approximants (\ref {eq:E0^(m)}) for $m\rightarrow \infty $. If the
square matrix in Eq.~(\ref{eq:E0^(m)}) is singular, the
corresponding approximant is omitted.

In principle, it may happen that $Z(t)=0$ for some values of $t$
in the complex $t$--plane. In such a case the cumulant expansion
(\ref{eq:t-exp}) converges for $t<|t_{s}|$, where $t_{s}$ is the
root of $Z(t)=0$ closest to the origin. (The location of such
singular points will obviously depend on the trial function
$\left|\varphi\right\rangle$). Therefore, matching the Taylor
series about $t=0$ for the two sides of equation
(\ref{eq:exp-exp}) may give rise to some difficulties, especially
because the expression in the right--hand side does not take into
account such singular points. Consequently, it is unclear that we
can successfully extrapolate the approximate expression for $E(t)$
thus derived to the limit $t\rightarrow \infty $. This aspect of
the problem was addressed by Witte\cite{W97} and Witte and
Shankar\cite{WS99} for some particular problems. In sections
\ref{sec:simple_models} and \ref {sec:oscillators} we will show
that under certain circumstances the main assumption of the CMX,
namely equation (\ref{eq:exp-exp}), does not apply to some simple
problems.

The exponential expansion (\ref{eq:exp-exp}) does in fact appear to take
into account the singular points of the inverted series\cite{SKMT97}. If we
keep the first two terms, solve for $t$ and differentiate with respect to $E$
we obtain
\begin{equation}
\frac{dt}{dE}=\frac{1}{b_{1}\left( E_{0}-E\right) }  \label{eq:dt/dE(app)}
\end{equation}
This approximate result partially agrees with the exact one that
we will derive for a two--level model in
section~\ref{sec:simple_models}.

\section{Simple models with $n$ states}

\label{sec:simple_models}

Some time ago Knowles\cite{K87} applied the CMX to the two--level model
\begin{equation}
\mathbf{H}=\left(
\begin{array}{ll}
0 & V \\
V & 1
\end{array}
\right)
\end{equation}
and concluded that the series (\ref{eq:E0^(m)}) for the reference function $%
\left| \varphi _{1}\right\rangle =\left(
\begin{array}{l}
1 \\
0
\end{array}
\right) $ converges to the exact eigenvalue for $|V|$ less than about $0.1$,
but as $|V|$ is increased the series converges to a result which is more and
more in error. He also found that the CMX series reproduces the correct $V$%
--power series for the same reference function but the series can become
violently oscillatory when the reference function is $\left| \varphi
_{2}\right\rangle =\frac{1}{\sqrt{2}}\left(
\begin{array}{l}
1 \\
1
\end{array}
\right) $. Besides, in the limit $V\rightarrow 0$ alternate
approximants for $\left| \varphi _{2}\right\rangle $ tend to zero
and infinity. Knowles\cite {K87} concluded that those results
placed some doubt on the claims that the series is convergent for
any reference function having a nonzero overlap with the exact
wavefunction.

Our numerical experiments for $V=0.1$ show that the CMX series (\ref
{eq:E0^(m)}) converges towards the ground--state energy $%
E_{0}=-0.009901951358$ when $\left| \varphi _{1}\right\rangle $ is the
reference function and to the excited state $E_{1}=1.009901951$ when the
trial vector is $\left| \varphi _{2}\right\rangle $. The CMX series also
converges for greater values of $V$; for example, for $V=1$ we obtain the
ground--state energy $E_{0}=-0.6180339887$ and the excited--state energy $%
E_{1}=1.618033988$ with the former and latter reference vectors,
respectively. The rate of convergence of the CMX series decreases
(increases) with $|V|$ in the former (latter) case. Convergence towards the
excited state is unexpected because there is no doubt that $E(t)$ always
tends to $E_{0}$ as $t\rightarrow \infty $ when $c_{0}\neq 0$. It may be for
this reason that convergence of the CMX series to excited states has been
ignored as far as we know. A reasonable explanation for such anomalous
behaviour is that the limit of the series of CMX approximants (\ref
{eq:E0^(m)}) is determined by the maximum overlap $\left| \left\langle \psi
_{j}\right. \left| \varphi \right\rangle \right| $ as suggested by the fact
that in the particular example just discussed $\left| \left\langle \psi
_{0}\right. \left| \varphi _{1}\right\rangle \right| >$ $\left| \left\langle
\psi _{1}\right. \left| \varphi _{1}\right\rangle \right| $ and $\left|
\left\langle \psi _{1}\right. \left| \varphi _{2}\right\rangle \right| >$ $%
\left| \left\langle \psi _{0}\right. \left| \varphi _{2}\right\rangle
\right| $ for all values of $V$. We will discuss this issue in more detail
below.

It is obvious that if the series of CMX approximants converges to an excited
state when $c_{0}\neq 0$ then the main assumption of the method, equation (%
\ref{eq:exp-exp}), is not valid in general. In what follows we explore this
point in more detail and write the CMX approximants to $E(t)$ as
\begin{equation}
E^{(M)}(t)=A_{0,M}+\sum_{j=1}^{M}A_{j,M}e^{-b_{j,M}t}  \label{eq:E^(M)(t)}
\end{equation}

The two--level system is suitable for a discussion of the convergence
properties of the $t$--expansion. In this case, equation (\ref{eq:E(t)_cj})
becomes
\begin{equation}
E(t)=\frac{E_{0}+\xi E_{1}e^{-t\Delta E}}{1+\xi e^{-t\Delta E}}
\label{eq:E(t)_cj_2x2}
\end{equation}
where $\Delta E=E_{1}-E_{0}$ and $\xi =|c_{1}|^{2}/|c_{0}|^{2}$. It is clear
that the expansion (\ref{eq:t-exp}) converges for all $|t|<|t_{s}|$, where
\begin{equation}
t_{s}=\frac{1}{\Delta E}\left( \ln \xi \pm \pi i\right)  \label{eq:ts}
\end{equation}
denotes the two singular points of $E(t)$ closest to the origin of the
complex $t$--plane. We see that the smallest radius of convergence occurs
when $\xi =1$.

For large $t$ we have the exponential expansion
\begin{equation}
E(t)=E_{0}+\Delta Eu\sum_{j=0}^{\infty }(-1)^{j}u^{j},\;u=\xi e^{-t\Delta E}
\end{equation}
and not the algebraic terms found by Witte\cite{W97} and Witte and Shankar%
\cite{WS99} for more elaborate models in the thermodynamic limit.
This exponential expansion converges for $u<1$. Although the
large--$t$ expansion of $E(t)$ exhibits the correct exponential
behaviour, it is not valid at the matching point $t=0$ unless $\xi
<1$. Therefore, we expect that the CMX exponential expansion
(\ref{eq:E^(M)(t)}) is valid only if $\xi <1$ ($|c_{0}|>|c_{1}|$).

If we solve equation (\ref{eq:E(t)_cj_2x2}) for $t$ and differentiate the
result we obtain
\begin{equation}
\frac{dt}{dE}=\frac{1}{\left( E_{0}-E\right) \left( E_{1}-E\right) }
\label{eq:dt/dE(2x2)}
\end{equation}
that resembles the approximate expression (\ref{eq:dt/dE(app)})
derived by \v{S}amaj et al\cite{SKMT97}. By means of this simple
model we realize why Stubbins\cite{S88} obtained reasonable
results from a Pad\'{e} analysis of the derivative $dt/dE$. Note
that in this case $t(E)$ exhibits logarithmic singularities
instead of the branch cut singularity appearing in the spin--1/2
isotropic antiferromagnetic XY chain\cite{W97}. Besides, the
derivative (\ref{eq:dt/dE(2x2)}) exhibits two singular points and
the expansion in powers of $E-E_{0}$\cite{TS11} converges only for
$|E-E_{0}|<E_{1}-E_{0}$.

In order to test the performance of the CMX we choose the even simpler
problem given by the diagonal Hamiltonian matrix
\begin{equation}
\mathbf{H}=\left(
\begin{array}{ll}
1 & 0 \\
0 & 2
\end{array}
\right)  \label{eq:H[1,0;0,2]}
\end{equation}
and an arbitrary reference state $\left| \varphi \right\rangle =\left(
\begin{array}{c}
(1+\xi )^{-1/2} \\
\xi ^{1/2}(1+\xi )^{-1/2}
\end{array}
\right) $, where $0\leq \xi <\infty $. The CMX expansion coefficients for $%
M=2$ are
\begin{eqnarray}
A_{0,2} &=&\frac{2\xi ^{3}+1}{\left( \xi +1\right) \left( \xi ^{2}-\xi
+1\right) }  \nonumber \\
A_{1,2} &=&\frac{\xi \left[ \left( \xi ^{2}-4\xi +1\right) \sqrt{\xi
^{2}-10\xi +1}-\xi ^{3}+11\xi ^{2}-11\xi +1\right] }{2\left( \xi +1\right)
\left( \xi ^{2}-\xi +1\right) \left( \xi ^{2}-10\xi +1\right) }  \nonumber \\
A_{2,2} &=&\frac{-\xi \left[ \left( \xi ^{2}-4\xi +1\right) \sqrt{\xi
^{2}-10\xi +1}+\xi ^{3}-11\xi ^{2}+11\xi -1\right] }{2\left( \xi +1\right)
\left( \xi ^{4}-11\xi ^{3}+12\xi ^{2}-11\xi +1\right) }  \nonumber \\
b_{1,2} &=&-\frac{\sqrt{\xi ^{2}-10\xi +1}+3\left( \xi -1\right) }{2\left(
\xi +1\right) }  \nonumber \\
b_{2,2} &=&\frac{\sqrt{\xi ^{2}-10\xi +1}-3\left( \xi -1\right) }{2\left(
\xi +1\right) }  \label{eq:A,b,M=2}
\end{eqnarray}
We appreciate that $A_{0,2}$ is closer to $E_{0}=1$ when $\xi <1$ and to $%
E_{1}=2$ when $\xi >1$ as shown by the expansions
\begin{eqnarray}
A_{0,2} &=&1+\xi ^{3}-\xi ^{6}+\xi ^{9}+\ldots ,\;\xi <1  \nonumber \\
A_{0,2} &=&2-\frac{1}{\xi ^{3}}+\frac{1}{\xi ^{6}}-\frac{1}{\xi ^{9}}+\ldots
,\;\xi >1
\end{eqnarray}

\begin{figure}[H]
\begin{center}
\bigskip\bigskip\bigskip \includegraphics[width=9cm]{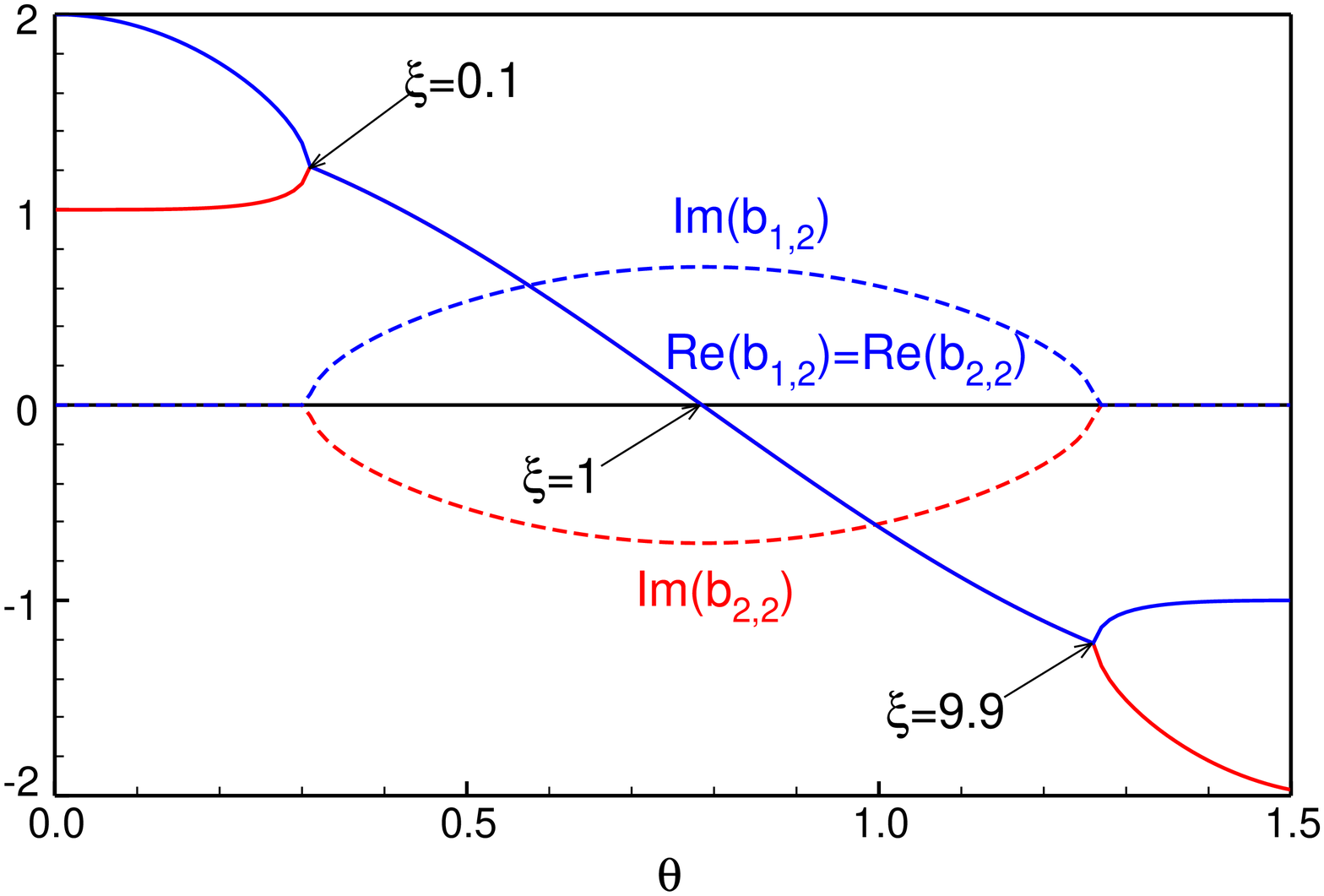}
\end{center}
\caption{Exponents $b_1$ and $b_2$ as functions of $\theta$ $\left(\xi=\frac{%
\sin(\theta)^2}{\cos(\theta)^2}\right)$}
\label{Fig:BM2}
\end{figure}

When $0<\xi <5-2\sqrt{6}\approx 0.10$ the nonlinear parameters
$b_{j,2}$ are real and positive. The CMX converges towards the
ground--state energy and $E^{(2)}(t)$ exhibits the correct
behaviour. When $5-2\sqrt{6}<\xi <1$ those nonlinear parameters
are complex conjugate each other $b_{1,2}=b_{2,2}^{*}$ with
${\mathrm Re}(b_{1,2})>0$. In this case the CMX still converges
towards $E_{0}$ and $E^{(2)}(t)$ is an almost reasonable
approximation to $E(t)$: it exhibits a minimum at some $t>0$ and
tends to a limit close to $E_{0}$ as $t\rightarrow \infty $. When
$1<\xi <5+2\sqrt{6}\approx 9.90$ the nonlinear parameters are
still complex conjugate each other $b_{1,2}=b_{2,2}^{*}$ but
${\mathrm Re}(b_{1,2})<0$. In this case the CMX converges towards
the excited--state energy and $E^{(2)}(t)$ exhibits a minimum at
some $t<0$ and tends to a
limit close to $E_{1}$ as $t\rightarrow -\infty $. Finally, when $\xi >5+2%
\sqrt{6}$ the nonlinear parameters $b_{j,2}$ are real and negative. The CMX
converges towards $E_{1}$ and $E^{(2)}(t)$ tends to a limit close to $E_{1}$
as $t\rightarrow -\infty $. This discussion is illustrated in Fig.~\ref
{Fig:BM2} that shows the real and imaginary parts of $b_{j,2}$ as functions
of $\theta, $ where $\xi =\frac{\sin (\theta )^{2}}{\cos (\theta )^{2}}$, $%
0<\theta <\pi /2$, that is a more convenient variable for plotting the
exponential parameters.

\begin{figure}[H]
\begin{center}
\bigskip\bigskip\bigskip \includegraphics[width=9cm]{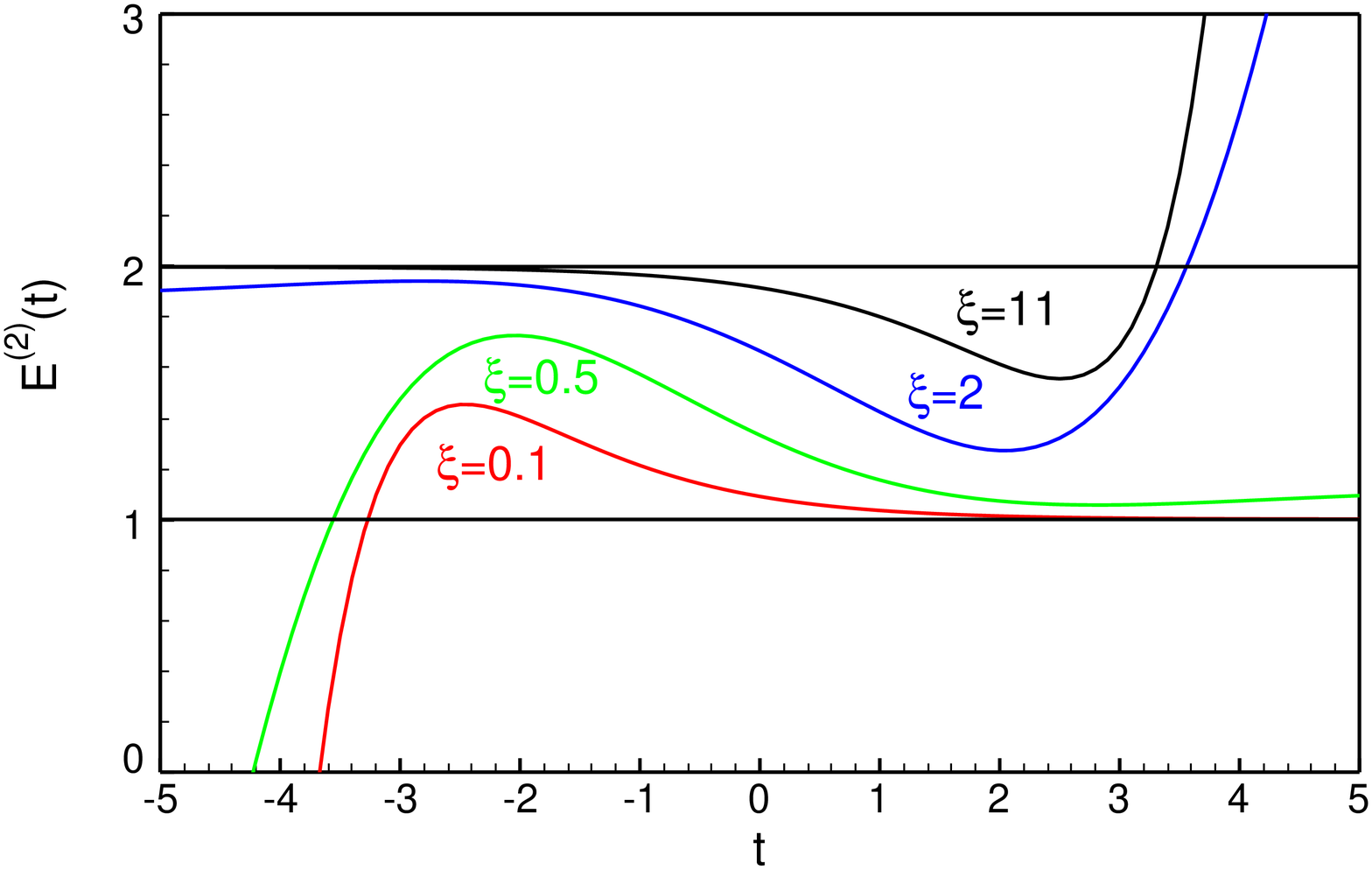}
\end{center}
\caption{Approximate generating functions $E^{(2)}$ for four values of $\xi$}
\label{Fig:E(2)}
\end{figure}

As a further illustration of the discussion above in what follows we show
the approximate function $E^{(2)}(t)$ in the four regions already indicated.
For example, for $\xi =0.1$

\[
E^{(2)}(t)=\frac{1002}{1001}+\frac{50}{143}e^{\frac{-13t}{11}}-\frac{20}{77}
e^{\frac{-14t}{11}}
\]
is a monotonous decreasing function that leads to a limit close to $E_{0}=1$
when $t\rightarrow \infty $. When $\xi =0.5$

\[
E^{(2)}(t)=\frac{10}{9}+\left[ \frac{2}{9}\cos {\left( \frac{\sqrt{15}t}{6}%
\right) }-\frac{2\sqrt{15}}{45}\sin {\left( \frac{\sqrt{15}t}{6}\right) }%
\right] e^{\frac{-t}{2}}
\]
leads to $10/9\approx 1$ as $t\rightarrow \infty $ but it exhibits a minimum
at $t\approx 2.8$. When $\xi =1$

\begin{equation}
E^{(2)}(t)=\frac{3}{2}-\frac{\sqrt{2}\sin {\left( \frac{\sqrt{2}t}{2}\right)
}}{4}
\end{equation}
oscillates and does not tend to any limit. The CMX approximants
$A_{0,M}$ oscillate yielding either $3/2$ or $\pm \infty $ when
$M$ is even or odd, respectively. When $\xi =2$ the approximate
generating function

\[
E^{(2)}(t)=\frac{17}{9}-\left[ \frac{2}{9}\cos {\left( \frac{\sqrt{15}t}{6}%
\right) }+\frac{2\sqrt{15}}{45}\sin {\left( \frac{\sqrt{15}t}{6}\right) }%
\right] e^{\frac{t}{2}}
\]
is unsuitable for $t>0$ and approaches the excited state as it tends to $%
17/9\approx 2$ as $t\rightarrow -\infty $. Finally, when $\xi =11$

\[
E^{(2)}(t)=\frac{2663}{1332}+\left( \frac{143\sqrt{3}}{2664}-\frac{55}{1332}%
\right) e^{t\left( \frac{\sqrt{3}}{12}+\frac{5}{4}\right) }-\left( \frac{143%
\sqrt{3}}{2664}+\frac{55}{1332}\right) e^{t\left( \frac{5}{4}-\frac{\sqrt{3}%
}{12}\right) }
\]
approaches the excited state as $t\rightarrow \infty $ as it tends to $%
2663/1332\approx 2$. Fig.~\ref{Fig:E(2)} illustrates the behaviours of those
functions.

Table~\ref{tab:A_0M_2x2} clearly shows that when $\xi =4$ $A_{0,M}$ already
converges towards $E_{1}=2$ as $M$ increases as discussed above. We conclude
that when $\xi >1$ the approximate functions $E^{(M)}(t)$ obtained by
matching the Taylor series about $t=0$ cannot exhibit the correct
exponential behaviour for $t\rightarrow \infty $. It is not difficult to
understand what happens when $\xi >1$ in the case of the two--level model.
First of all note that $\lim_{t\rightarrow -\infty }E(t)=E_{1}$. Second, the
exponential expansion of the cumulant--generating function (\ref
{eq:E(t)_cj_2x2})
\begin{equation}
E(t)=E_{1}-\Delta Eu^{-1}\sum_{j=0}^{\infty }(-1)^{j}u^{-j}
\label{eq:2x2_exp_exp_2}
\end{equation}
converges for all $t \leq 0$ when $\xi >1$. Third, the match between the
exponential expansion (\ref{eq:E^(M)(t)}) and the $t-$expansion at $t=0$ is
valid for both, positive and negative, values of $t$. Therefore, when $%
|c_{1}|>|c_{0}|$ the CMX simply chooses the convergent exponential expansion
for $t<0$ and the result is the energy of the excited state.

The CMX fails completely when $\xi =1$ because both exponential expansions ($%
t>0$ and $t<0$) are divergent at $t=0$. It is worth noting that the original
approach of Horn an Weinstein\cite{HW84} in terms of Pad\'{e} approximants $%
[M/M](t)$ for $E(t)$ yields accurate results in this most unfavourable case.
These approximants exhibit either a saddle point or a minimum when $M$ is
either odd or even, respectively. From such stationary points we obtain $%
E_{0}=0.9993674107,1.000359205,0.9999637967,1.000025380$ with $M=5,6,7,8$,
respectively. In addition to it, these Pad\'{e} approximants yield
increasingly accurate estimates of the poles of $E(t)$ at $t_{s}=\pm
(2j+1)\pi i$, $j=0,1,\ldots $.

\begin{figure}[H]
\begin{center}
\bigskip\bigskip\bigskip \includegraphics[width=9cm]{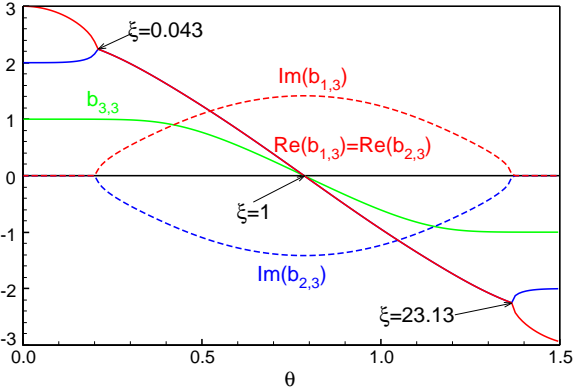}
\end{center}
\caption{Exponents $b_{j,3}$ as functions of $\theta$ $\left(\xi=\frac{%
\sin(\theta)^2}{\cos(\theta)^2}\right)$}
\label{Fig:BM3}
\end{figure}

The behaviour of the exponential parameters discussed above is not
restricted to the case $M=2$. Fig.~\ref{Fig:BM3} shows that two of the
exponential parameters $b_{j,3}$ behave in a similar way. The third one,
which we arbitrarily chose to be $b_{3,3}$, is real because the three
parameters are solutions to a cubic equation. In fact, the exponential
parameters $b_{j,M}$ are in general the $M$ roots of the pseudo secular
determinant
\begin{equation}
\left| I_{i+j+1}-bI_{i+j}\right| _{i,j=1}^{M}=0  \label{eq:secular_Iij}
\end{equation}

\begin{figure}[H]
\begin{center}
\par
\includegraphics[width=9cm]{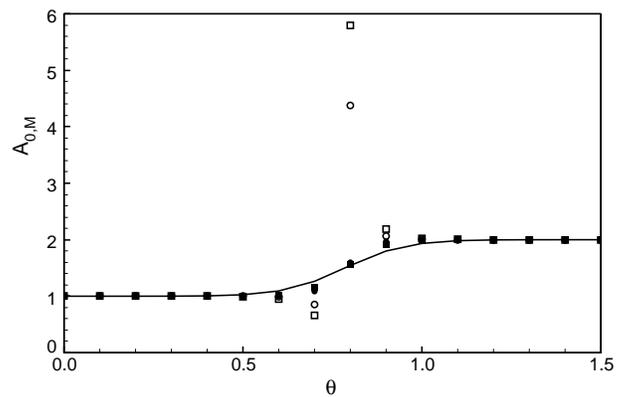}
\end{center}
\caption{Approximants $A_{0,M}$ with $M=2$ (line), $M=3$ (squares), $M=4$
(filled squares), $M=5$ (circles) and $M=6$ (filled circles) as a function
of $\theta$ }
\label{fig:A0M(theta)}
\end{figure}

Fig.~\ref{fig:A0M(theta)} shows $A_{0,M}(\theta )$ for some values
of $M$. We appreciate that the series of CMX approximants
converges towards $E_{0}=1$ when $\theta <\pi /4$ and towards
$E_{1}=2$ when $\theta >\pi /4$ and diverges when $\theta $ is
close to $\pi /4$ ($\xi =1$). It is a further illustration and
confirmation of the arguments above.

These examples clearly show that the condition $c_{0}\neq 0$ is
insufficient to guarantee the validity of equation
(\ref{eq:exp-exp}) that is the main assumption of the CMX. In
fact, they suggest that if $|c_{0}|\neq 0$ is not the largest
overlap, then the CMX series may converge towards an excited state
in which case $E^{(M)}(t)$ will not be an acceptable approximation
to $ E(t)$ for $t>0$, except for sufficiently small values of $t$.
Besides, for some reference states (like those with $\xi =1$ for
the two--level model) the CMX does not converge at all.

It is not difficult to show that the results for the two--level model also
apply to arbitrary problems. For example, if we choose the reference state
\begin{equation}
\left| \varphi \right\rangle =c_{0}\left| \psi _{0}\right\rangle
+c_{j}\left| \psi _{j}\right\rangle
\end{equation}
then we draw conclusions similar to those above for $\xi
=|c_{j}|^{2}/|c_{0}|^{2}$ because we already have a two--level
problem.

We have also carried out a numerical and analytical study of the CMX for
Hamiltonian matrices of greater dimension. For example, if we choose a
diagonal matrix with elements $H_{ij}=j\delta _{ij}$, $i,j=0,1,\ldots ,9$
and the reference vector with coefficients $c_{j}=(1+\delta _{j0})/\sqrt{13}$%
, $j=0,1,\ldots ,9$ then the generating function $E(t)$ is a ratio of
polynomial functions of $u=e^{-t}$. The $u$--power series (large--$t$
expansion) converges for all $|u|<1.074531738$ that is quite close to unity.
For this reason we expect some difficulties when matching the $t$-- and $u$%
--series. In fact, in this case the CMX series does not appear to converge
towards the ground--state energy (or the convergence rate is too small for
practical purposes).

\section{Oscillators}

\label{sec:oscillators}

The conclusions of the preceding section are not restricted to the
particular case in which only a finite number of states contribute
to $E(t)$. In what follows we consider quantum--mechanical systems
with an infinite number of bound states. In order to keep present
discussion as simple as possible in what follows we consider the
harmonic oscillator
\begin{equation}
\hat{H}=-\frac{d^{2}}{dx^{2}}+x^{2}
\end{equation}
If we choose
\begin{equation}
\left| \varphi \right\rangle =N\exp \left( -2x^{2}/5\right)
\end{equation}
where $N$ is a normalization factor, then the CMX converges towards the
ground state because $|c_{0}|>|c_{j}|$ for all $j$. In this case the
exponential parameters for the case $M=3$ are $b_{1,3}\approx 4$, $%
b_{2,3}\approx 8$ and $b_{3,3}\approx 12.6$ that are quite close to the
actual excitations $\Delta E_{j}=E_{2j}-E_{0}$ as expected.

If, on the other hand, we choose
\begin{equation}
\left| \varphi \right\rangle =N\left( x^{2}-\frac{1}{2}\right) \exp \left(
-2x^{2}/5\right)
\end{equation}
then the CMX converges towards the second excited state because $%
|c_{2}|>|c_{j}|$. Note that in this case $\left\langle \psi _{0}\right.
\left| \varphi \right\rangle \neq 0$ and $\lim_{t\rightarrow \infty }E(t)=1$
; consequently $E^{(M)}(t)$ will not be an acceptable approximation to $E(t)$
beyond a sufficiently small neighbourhood of $t=0$. This conclusion is
consistent with the exponential parameters $b_{1,3}\approx -3.87$, $%
b_{2,3}\approx 4$ and $b_{3,3}\approx 9.29$.

If we choose
\begin{equation}
\left| \varphi \right\rangle =\frac{1}{\sqrt{2}}\left( \left| \psi
_{0}\right\rangle +\left| \psi _{2}\right\rangle \right)
\end{equation}
then the CMX series $A_{0,M}$ alternate between $\pm \infty $ and
$3$ for $M$ odd and even, respectively. This result is not
surprising in the light of our previous discussion of two--level
models. At first sight one may think that it suggests that the CMX
fails when two overlaps are equal $|c_{i}|=|c_{j}|$ and greater
than any other $|c_{k}|$; however, this is not the case. For
example, the reference function
\begin{equation}
\left| \varphi \right\rangle =\frac{\sqrt{\frac{1040\sqrt{2}}{4217}-\frac{64%
}{4217}}}{\pi ^{\frac{1}{4}}}e^{-x^{2}}\left( x^{2}\left( 3\sqrt{2}+1\right)
-\sqrt{2}+1\right)
\end{equation}
satisfies $c_{0}=c_{2}=\sqrt{\frac{33280\sqrt{2}}{113859}-\frac{2048}{113859}%
}$ but the CMX converges towards the ground state. In this case we obtain $%
b_{1,3}=b_{2,3}^{*}\approx 8.53+1.76i$ and $b_{3,3}\approx 4.06.$

As a nontrivial example we choose the anharmonic oscillator
\begin{equation}
-\frac{d^{2}}{dx^{2}}+x^{4}  \label{eq:AHO}
\end{equation}
and the reference states
\begin{eqnarray}
\left| \varphi \right\rangle _{g} &=&N\exp \left( -3x^{2}/2\right)  \nonumber
\\
\left| \varphi \right\rangle _{e} &=&N\left( x^{2}-\frac{1}{4}\right) \exp
\left( -3x^{2}/2\right)
\end{eqnarray}
The CMX series converges towards the ground--state energy in the former case
($b_{1,3}\approx 6.34$, $b_{2,3}\approx 17.19$, $b_{3,3}\approx 34.42$) and
oscillates around the second excited--state energy in the latter one ($%
b_{1,3}\approx -2.61$, $b_{2,3}\approx 9.04$, $b_{3,3}\approx 26.1$) as
shown in Table~\ref{tab:A_0M_AHO}. One expects that $|c_{0g}|>|c_{jg}|$ for
all $j>0$ and that $|c_{2e}|>|c_{je}|$ for all $j\neq 2$. What we know for
sure is that $E^{(M)}(t)$ will not be an acceptable approximation to $E(t)$
when $\left| \varphi \right\rangle =\left| \varphi \right\rangle _{e}$
except in a neighbourhood of $t=0$ as indicated by the exponential
parameters $b_{j,3}$.

\section{Conclusions}

\label{sec:conclusions}

The main assumption of the CMX is that the exponential expansion (\ref
{eq:exp-exp}) is a suitable approximation to the cumulant--generating
function (\ref{eq:E(t)}) when $|c_{0}|\neq 0$. Up to now, it was believed
that the only limitations of the method were singularities in the CMX
approximants (\ref{eq:E0^(m)}) or an occasional divergence for some
reference state. By means of simple quantum--mechanical models we have shown
that the performance of the CMX is mainly determined by the overlaps $%
|c_{j}| $. Present investigation strongly suggests that the series of the
CMX approximants (\ref{eq:E0^(m)}) converges towards the energy of the state
with greatest overlap with the reference function. Thus, we have the
following situations:

\begin{itemize}
\item  When $|c_{0}|>|c_{j}|$ for all $j>0$ the series $A_{0,M}$ converges
towards the ground--state energy and we can say that the CMX is valid.

\item  The series $A_{0,M}$ converges towards the energy of an excited state
(or oscillates about it). In such a case we may say that the CMX is still
useful but its main assumption embodied in equation (\ref{eq:exp-exp}) is no
longer valid. This situation may happen when the reference state exhibits
the largest overlap with an excited state.

\item  The series $A_{0,M}$ does not converge at all and the CMX is of no
practical utility.
\end{itemize}

In no case does the CMX converge to a meaningless limit as
suggested by Knowles\cite{K87} based on results for a two--level
model. The main conclusions of this paper agree with those drawn
earlier that the CMX and its variants should be applied with
extreme care in order to obtain useful results\cite
{K87,MPM91,LL93,MZM94,MZMMP94,U95,AF09b,AFR10}. In fact, in most
cases the Rayleigh--Ritz variational method in the Krylov space is
by far preferable to the CMX and its variants\cite{AF09b,AFR10}.
We have also shown that the original method of Horn and
Weinstein\cite{HW84} yields accurate results in a case in which
the CMX badly fails. The choice of an appropriate trial state may
improve the results dramatically as illustrated by a recent study
on the Rabi Hamiltonian\cite{TS11}.

Finally, one of the main contributions of this paper is that the analysis of
the exponential parameters $b_{j,M}$ may provide an indication on the rate
of convergence of the CMX approximants. If all such parameters are real and
positive one expects a reasonable convergence towards the ground state
(provided that $c_{0}\neq 0$). This analysis is greatly facilitated by the
fact that those exponential parameters are the roots of the relatively
simple pseudo secular equation (\ref{eq:secular_Iij}).

\begin{table}[H]
\caption{Convergence of the CMX towards the ground--state energy (second
column) and excited--state energy (third column) of the two--level model (
\ref{eq:H[1,0;0,2]}) for $\xi=1/4$ and $\xi=4$, respectively.}
\label{tab:A_0M_2x2}
\begin{center}
\par
\begin{tabular}{rll}
\hline
$M$ & $A_{0,M}$ & $A_{0,M}$ \\ \hline
2 & 1.015384615 & 1.984615384 \\
4 & 1.000975609 & 1.99902439 \\
6 & 1.000061031 & 1.999938968 \\
8 & 1.000003814 & 1.999996185 \\
10 & 1.000000238 & 1.999999761 \\
12 & 1.000000000 & 2.000000000 \\ \hline
\end{tabular}
\par
\end{center}
\end{table}

\begin{table}[H]
\caption{Convergence of the CMX towards the ground--state $(g)$ and
second--excited state $(e)$ energies of the anharmonic oscillator (\ref
{eq:AHO})}
\label{tab:A_0M_AHO}
\begin{center}
\par
\begin{tabular}{rll}
\hline
$M$ & $A_{0,M}(g)$ & $A_{0,M}(e)$ \\ \hline
5 & 1.060692159 & 7.439371257 \\
10 & 1.060363186 & 7.456069907 \\
15 & 1.060362073 & 7.450017954 \\
20 & 1.060362093 & 7.451366303 \\
25 & 1.060362090 & 7.455118704 \\
30 & 1.060362090 & 7.454183973 \\
35 & " & 7.451642486 \\
40 & " & 7.454364274 \\
50 & " & 7.454214745 \\
60 & " & 7.453864737 \\
70 & " & 7.455066766 \\
80 & " & 7.455185890 \\
90 & " & 7.453941990 \\
100 & " & 7.453833053 \\ \hline
RPM\cite{FMT89a,FMT89b} & 1.060362090 & 7.455697938 \\ \hline
\end{tabular}
\par
\end{center}
\end{table}

\end{document}